\documentclass[useAMS,usenatbib,usegraphicx,usedcolumn]{mn2e}
\usepackage{times,amssymb}
\title[HDF\,130 and 6C\,0905+3955]{The inverse-Compton ghost HDF\,130 and the giant radio galaxy 6C\,0905+3955: matching an analytic model for double radio source evolution}
\author[P. Mocz et. al.]{P. Mocz$^{1,2}$\thanks{E-mail: pmocz@fas.harvard.edu (PM); acf@ast.cam.ac.uk (ACF); kmb@astro.ox.ac.uk (KMB); ptg@astro.ox.ac.uk (PTG); schapman@ast.cam.ac.uk (SCC); djs@ncra.tifr.res.in (DJS)},
 A.C. Fabian$^{2}$\footnotemark[1], Katherine M. Blundell$^{3}$\footnotemark[1],  P.T. Goodall$^{3}$\footnotemark[1], \newauthor S.C. Chapman$^{2}$\footnotemark[1] and D.J. Saikia$^{4}$\footnotemark[1]\\
$^{1}$Harvard University, Cambridge, MA 02138, USA\\
$^{2}$Institute of Astronomy, Madingley Road, Cambridge CB3 0HA\\
$^{3}$Astrophysics, University of Oxford, Keble Road, Oxford OX1 3RH\\
$^{4}$National Centre for Radio Astrophysics, TIFR, Pune University Campus, Post Bag 3, Pune 411 007, India}
\begin{document}

\date{subm. to MNRAS, 22 August 2010, accepted for publication}

\pagerange{\pageref{firstpage}--\pageref{lastpage}} \pubyear{2011}

\maketitle

\label{firstpage}

\begin{abstract}
We present new GMRT observations of HDF\,130, an inverse-Compton (IC) ghost of a giant radio source that is no longer being powered by jets. We compare the properties of HDF\,130 with the new and important constraint of the upper limit of the radio flux density at $240$~MHz to an analytic model. We learn what values of physical parameters in the model for the dynamics and evolution of the radio luminosity and X-ray luminosity (due to IC scattering of the cosmic microwave background (CMB)) of a Fanaroff-Riley II (FR~II) source 
are able to describe a source with features (lobe length, axial ratio, X-ray luminosity, photon index and upper limit of radio luminosity) similar to the observations. HDF\,130 is found to agree with the interpretation that it is an IC ghost of a powerful double-lobed radio source, and we are observing it at least a few Myr after jet activity (which lasted $5$--$100$~Myr) has ceased. The minimum Lorentz factor of injected particles into the lobes from the hotspot is preferred to be $\gamma\sim10^3$ for the model to describe the observed quantities well, assuming that the magnetic energy density, electron energy density, and lobe pressure at time of injection into the lobe are linked by constant factors according to a minimum energy argument, so that the minimum Lorentz factor is constrained by the lobe pressure. We also apply the model to match the features of 6C\,0905+3955, a classical double FR~II galaxy thought to have a low-energy cutoff of $\gamma\sim10^4$ in the hotspot due to a lack of hotspot inverse-Compton X-ray emission. The models suggest that the low-energy cutoff in the hotspots of 6C\,0905+3955 is $\gamma\gtrsim 10^3$, just slightly above the particles required for X-ray emission.
\end{abstract}


\section{Introduction}\label{Introduction}

In this work we establish whether a consistent interpretation can be found for the currently observed properties of the double-lobed sources HDF\,130 and 6C\,0905+39, the former of which is thought to no longer have current jet activity. We identify at which stage in their life cycles HDF\,130 and 6C\,0905+3955 could be by comparing their observable features (lobe length, axial ratio, X-ray luminosity, photon index, radio flux density limit) to an analytic model for the dynamics and evolution of X-ray and radio emission of an active FR~II object whose jets switch off after a time $t_{\rm j}$ developed in \citet{2011MNRAS.413.1107M}. The models also help suggest what the physical parameters of the sources such as jet energy, jet lifetime, ambient density parameters, and injection spectrum parameters may be, some of which are difficult to determine from observation alone. 

{\it Hubble Deep Field}~(HDF)~130 is an extended X-ray source observed in the {\it Chandra Deep Field}-North X-ray image \citep{2003AJ....126..539A}. All six of the extended  X-ray sources found in the $1$~Ms exposure were attributed to clusters and groups by \cite{2002AJ....123.1163B}. 
However, HDF\,130 has since been realised to be a double-lobed structure with extended X-ray emission due to IC scattering of the CMB \citep{2009MNRAS.395L..67F}.
While the jet is turned on in a powerful radio source, $\gamma\sim10^4$ electrons (assuming typical magnetic field strengths) required to generate GHz synchrotron radiation in the radio band lose their energy due to radiative losses much more quickly than the $\gamma\sim10^3$ electrons responsible for upscattering the CMB photons; these losses are compounded by the expansion of the plasma. Thus after jet activity ceases, the IC X-ray emission lasts longer than radio emission and the source will appear as an IC ghost of a radio lobe for some period of time. On a morphological basis, and also given the non-thermal nature of the spectrum of the extended X-ray emission from HDF\,130 \citep{2009MNRAS.395L..67F}, the extended source is most likely the lobes of a formerly powerful Fanaroff-Riley II (FR~II) \cite{1974MNRAS.167P..31F} galaxy.

HDF\,130 is approximately $690$~kpc across as determined by \cite{2009MNRAS.395L..67F}, and such an extent is not exceptional (e.g. \citealt{2008MNRAS.390..595M}). The extended emission has a steep photon index of $\Gamma=2.65$, which could indicate significant synchrotron cooling. HDF\,130 is about half as bright at X-ray wavelengths as the giant powerful radio galaxy 6C\,0905+3955 \citep{2008MNRAS.386.1774E}, which has a spectral index of $\Gamma=1.61$, suggesting that HDF\,130 may be viewed at a later stage in its life cycle than 6C\,0905+3955.

6C\,0905+3955 is a powerful FR~II galaxy, approximately $945$~kpc in diameter \citep{2006ApJ...644L..13B}. The source characteristics mandate a low-energy cutoff of freshly injected particles in the hotspot above $\gamma\sim10^3$ due to the absence of X-ray emission from the hotspot (but not the lobes) \citep{2006ApJ...644L..13B,2008MNRAS.386.1774E}. The lobes, containing older plasma, do have $\gamma\sim10^3$ particles required for observing IC scattering on the CMB in the X-ray. The higher energy particles injected from the hotspot into the lobes undergo energy loss and hence result in the presence of plentiful $\gamma\sim10^3$ particles in the lobes.
Extended IC X-ray emission has been observed in other sources as well, such as 3C\,294 ($z=1.786$) \citep{2003MNRAS.341..729F}, 4C\,23.56 ($z=2.48$) \citep{2007MNRAS.376..151J} and 4C41.17 ($z=3.8$) \citep{2003ApJ...596..105S}. The CMB energy density is proportional to $(1+z)^4$, cancelling the dimming due to distance, and thus extended X-ray emission may be observable at both low and high redshifts \citep{1969Natur.221..924F}.

\section{Observations of HDF\,130 and 6C\,0905+39, and the model for evolution of an FR~II object}\label{obs}
\subsection{GMRT observations of HDF\,130 }\label{GMRT}
\begin{figure*}
\centering
\includegraphics[width=\textwidth]{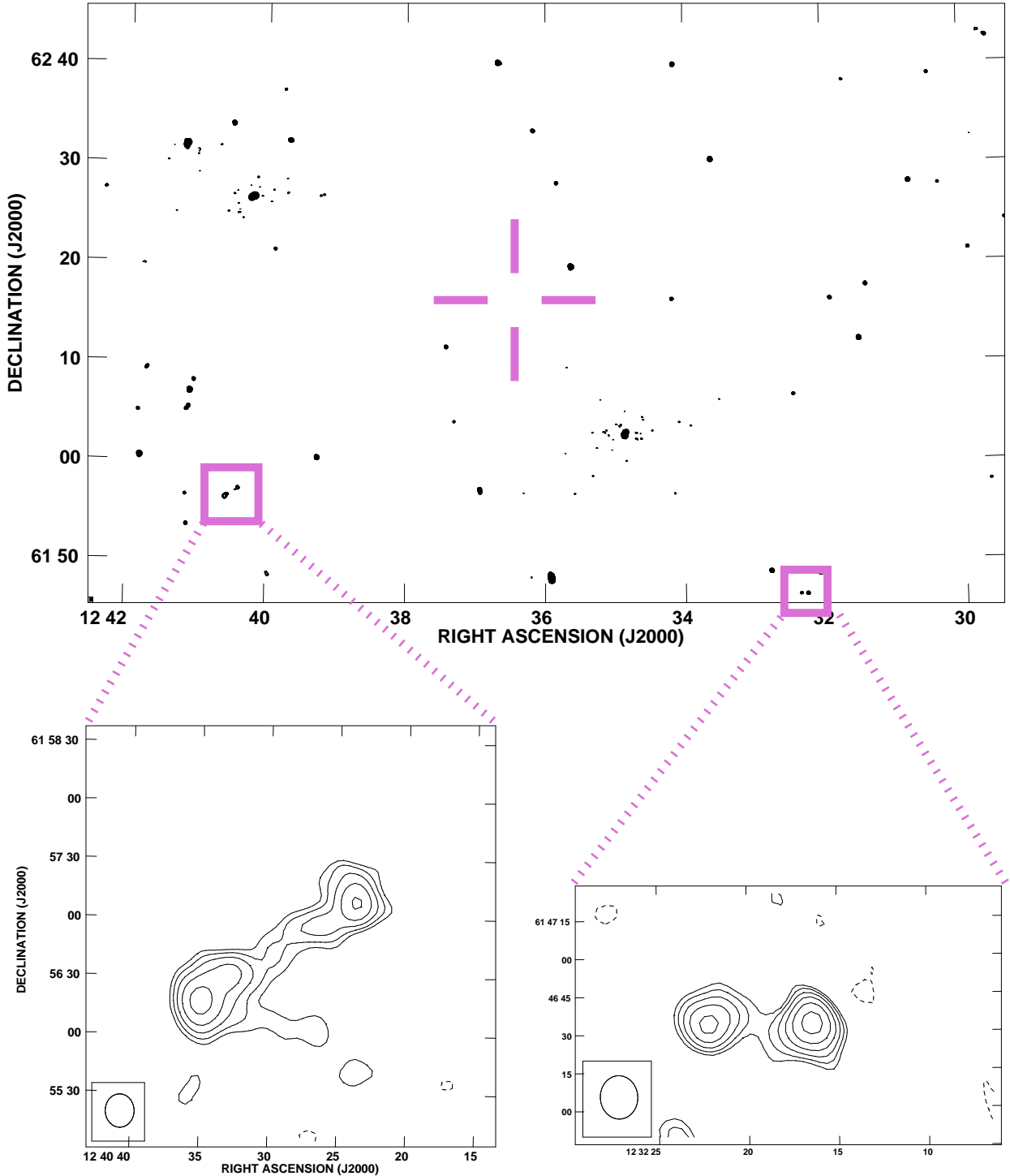}
\caption{Wide-field image of the region of sky surrounding HDF\,130, at $240$~MHz, observed by the GMRT (see \S~\ref{HDF130obs}). The lowest contour on the main image is $20$~mJy~beam$^{-1}$ while those in the insets are $4$~mJy~beam$^{-1}$. The insets show two double radio sources far from the phase centre.}
\label{fig:overlay}
\end{figure*}

We observed the target HDF\,130 for $9$ hours on 2008 Oct 25 using the Giant Metre-wave Radio Telescope (GMRT) at $240$~MHz. The observation was made in spectral line mode and had a total bandwidth of $8$~MHz, consisting of $128$ channels each of $62.5$~kHz, for each of RR and LL polarisations, which facilitated both high-fidelity imaging across the primary beam and also efficacious excision of radio-frequency interference. Absolute amplitudes were set using observations of 3C\,286. Observations of 1252+565, our chosen phase calibrator, were interleaved throughout the observation. After allowing for observations of the calibrator sources, the total on-target time was $7.3$~hours giving good UV-coverage to facilitate deconvolution. These data were reduced using standard wide-field procedures within AIPS, including facetting across the full primary beam. The innermost region of the resultant image is shown in Figure~\ref{fig:overlay}, which has resolution of $17$ arcsec by $14$~arcsec. The upper limit of the radio flux density of HDF\,130 at $240$~MHz in an $81$ arcsec by $15$ arcsec area centered on the source is $11$~mJy. The $3\sigma$ flux density upper limit is $33$~mJy.

\subsection{Previous HDF\,130 observations}\label{HDF130obs}
HDF\,130 was most recently analysed in the X-ray by \cite{2009MNRAS.395L..67F}. The source is a massive ($\sim5\times 10^{11}$~M$_{\odot}$ stellar population \citep{2009MNRAS.395.1249C}) elliptical galaxy at $z=1.99$ with a roughly double-lobed structure in the X-ray and a compact radio nucleus. Each lobe is approximately $345$~kpc long with an axial ratio (single lobe length divided by width of the lobe) of $2$ \citep{2009MNRAS.395L..67F}. The X-ray image of HDF\,130 observed in the {\it Chandra Deep Field}-North $2$~Ms exposure was best modelled with a photon index of $\Gamma=2.65$ by \cite{2009MNRAS.395L..67F} and has a $2$--$10$~keV luminosity of $5.4\times10^{43}$erg~s$^{-1}$.

\subsection{Previous 6C\,0905+3955 observations}

The powerful $z=1.88$ FR~II source 6C\,0905+3955 was most recently observed in the X-ray by {\it XMM-Newton} \citep{2008MNRAS.386.1774E}. The projected size of the source is $945$~kpc \citep{2006ApJ...644L..13B} with an axial ratio of $8$, although the source has arm-length asymmetry (the ratio of arm lengths is $1.6$) \citep{1995MNRAS.277..995L,2006ApJ...644L..13B}. The extended X-ray lobe emission was modelled with a photon index of $\Gamma=1.61$ and has a $2$--$10$~keV luminosity of $1.5\times 10^{44}$erg~s$^{-1}$ \citep{2008MNRAS.386.1774E}. The MERLIN $408$~MHz image gives a radio luminosity of the lobes of $8.4\times 10^{43}$erg~s$^{-1}$ \citep{1995MNRAS.277..995L}.

\subsection{Model for evolution of a double radio source}\label{evo}

We use a model for evolution of double radio sources developed in \citet{2011MNRAS.413.1107M}, where the full details may be found,  in order to determine the source properties of HDF\,130 and 6C\,0905+3955 and their evolutionary stage at the time of observation. 
Here we outline some of the basic features of the model.
The model is an analytic one for the dynamics and evolution of the radio luminosity and X-ray luminosity (due to IC scattering of the CMB) of FR~II radio galaxies. It accounts for injection of relativistic particles into the lobes of radio galaxies, and for adiabatic, synchrotron and IC energy losses to describe the evolution of the emission in the radio and the X-ray bands assuming a powerful double-lobed radio galaxy whose jets turn off after a typical jet lifetime. The model is based on the formalisms of \cite{1997MNRAS.292..723K}, \cite{1997MNRAS.286..215K}, \cite{1999AJ....117..677B} and \cite{2010MNRAS.407.1998N}.

The model is characterized by the jet power $Q_{\rm j}$ (per jet), jet lifetime $t_{\rm j}$, ambient density described by
\begin{equation}
\rho(r)=\rho_0(r/a_0)^{-\beta}
\label{eq:environment}
\end{equation}
and a power-law injection spectrum into the lobe of
\begin{equation}
n(\gamma_{\rm i},t_{\rm i})d\gamma_{\rm i}=n_0\gamma_{\rm i}^{-p}d\gamma_{\rm i}
\label{eq:ni}
\end{equation}
with $\gamma_{\rm i}$ between $\gamma_{\rm min}$ and $\gamma_{\rm max}$.

In the model, we have injection of relativistic electrons into the expanding lobe until jet activity stops and no further particles are added into the lobe.
The energy loss equation describes the time evolution of the Lorentz factors of the electrons:
\begin{equation}
\frac{d\gamma}{dt}=-\gamma\frac{1}{3}\frac{1}{V_{\rm l}}\frac{dV_{\rm l}}{dt}-\frac{4}{3}\frac{\sigma_{\rm T}}{m_{\rm e} c}\gamma^2 (u_{\rm B}+u_{\rm c})
\label{eq:loss}
\end{equation}
where the first term is the energy loss due to the adiabatic expansion of the lobe of volume $V_{\rm l}$ and the second term describes the synchrotron and IC losses. Here $m_{\rm e}$ is the mass of an electron, $u_{\rm c}=a(2.7~{\rm K}~(1+z))^4$ is the CMB photon energy density at the redshift of the source and $a=7.565\times10^{-16}~{\rm J}~{\rm K}^{-4}~{\rm m}^{-3}$ is the radiation constant.
Once no more fresh particles are injected into the lobes, the X-ray photon index will steepen from what is expected from the injection spectrum index due to the synchrotron and IC energy losses.

The pressure in the lobes, the energy density of the electrons, and the energy density of the magnetic field in the model are related by constants of order unity based on minimum energy arguments, adapted from \cite{1997MNRAS.292..723K}. The strength of the magnetic field is governed by the jet power (higher jet power corresponding to higher magnetic fields), and the magnetic fields do decrease in the evolution of the source as the lobes expand. The magnetic fields range between $0.1$--$10$~$\mu$G.

Typical jet energies may range from $5\times 10^{37}$~W to $10^{40}$~W, and jet lifetimes may vary from $10^6$~yr to $10^9$~yr.
More powerful jets give rise to lobes that are brighter and grow larger. During the time the jet is on, the radio luminosity decreases with time but the X-ray luminosity increases because higher energy electrons, losing their energy more quickly, 
downshift to Lorentz factors of $\gamma\sim 10^3$ needed for the IC scattering of CMB photons to $1$~keV energies and the injection of new energetic electrons into the lobe by the jet more than compensates for the original $\gamma\sim 10^3$ electrons losing energy. 
Once the jet is off, there is no injection of new particles and the radio and X-ray luminosities fall (the fall is faster in the radio as higher energy electrons required for synchrotron radiation lose energy more rapidly).

One set of environmental parameters inferred from observations is: $\beta=1.5$, $a_0=10~{\rm kpc}$ and $\rho_0=1.67\times 10^{-23}~{\rm kg}~{\rm m}^{-3}$ \citep{1999AJ....117..677B}. 
A less dense environment would allow for sources
to grow larger but luminosity then falls more quickly due to increased adiabatic losses.
The injection index $p$ is between $2$ and $3$ \citep{1987MNRAS.225....1A}. The maximum injected Lorentz factor $\gamma_{\rm max}$ is set to $10^6$ as Lorentz factors of $\gamma\sim10^3$ are required to produce upscattering of the CMB in the X-ray and Lorentz factors of $\gamma\geq 10^4$ are needed for GHz synchrotron radiation in the radio for typical magnetic field strengths of $B\sim0.1$--$10$~$\mu$G. The minimum injected Lorentz factor $\gamma_{\rm min}$ may in principle be as low as $1$.

The output luminosities of the model are in units of power per frequency per steradian, which is converted to a power (units of erg~s$^{-1}$) by multiplying by $4\pi$ steradians and the frequency $\nu$ of the emission.

\begin{table}
\centering
\caption{Model parameters tested}\label{tblpar}
\begin{tabular}{@{}cc@{}}
\hline\hline
\noalign{\smallskip}
Parameter & values \\[2pt]
\hline 
\noalign{\smallskip}
$Q_{\rm j}$ (W) & $5\times10^{37}, 10^{38}, 2\times10^{38}, 5\times10^{38}, \ldots, 10^{40}$ \\
$t_{\rm j}$ (yr) & $10^5, 5\times 10^5, 10^6, 5\times 10^6, \ldots, 10^9$ \\
$\beta$ & $1.5,2$ \\
$\rho_0$~(kg m$^{-3}$) & $1.67\times 10^{-23}, 1.67\times 10^{-22}$ \\
$\gamma_{\rm min}$ & $1,10^3,10^4$  \\
$p$ & $2.14,2.5,3$ \\
\hline
\end{tabular}
\end{table}

\section{Results}\label{results}

Our physical model for the radio source has $8$ parameters (jet energy, jet lifetime, minimum injected Lorentz factor $\gamma_{\rm min}$, injection power spectrum index $p$, $\beta$ and $\rho_0$ characterizing the environment, and the time of observation). We have five predicted observables (lobe length, axial ratio, X-ray luminosity, photon index, radio flux density limit). Therefore we cannot fit the model to the $5$ data points since we would be overfitting. However, we can do something simple. 
We can consider a wide range of parameters which observations suggest describe most double-lobed radio sources
and test which combinations of parameters can yield an object that is similar to HDF\,130 or 6C\,0905+3955. Looking at the sets of 
parameters that do well will indicate what types of scenarios may be possible. We are not able to conclude the value of any physical parameter specifically but can learn that certain values for a specific parameter in the model may be unable to reproduce observed source properties. The length of time of the observational window during which a specific set of physical parameters leads to a model similar to the observed source gives an estimate of the likelihood of the model. Without considering the observational window, all the sets of parameters that describe a source similar to the observed properties are equally likely candidates to describe the source. But considering the observational window, a model that describes a source with similar features to the observed source for only a very brief period of time is less probable to be the correct descriptor than a model that does so for a longer window. In order to estimate whether one value, $x_1$ for a parameter $x$  may be preferred over another, $x_2$,
in a Bayesian sense,
one may compare the total length of time of the congruent observational windows of all the models investigated with $x=x_1$ to all the models with $x=x_2$.

The source parameters being fit are obtained from the observations mentioned in \S~\ref{obs}. We list the parameters here: 
HDF\,130 has lobe of length $345$~kpc, axial ratio $2.0$, $1$~keV X-ray luminosity of $8.5\times 10^{43}$~erg~s$^{-1}$, $\Gamma=2.65$ and upper limit of the radio luminosity at $240$~MHz of  $2.3\times 10^{42}$~erg~s$^{-1}$.
6C\,0905+3955 has lobe of length $472$~kpc, axial ratio $8.0$, $1$~keV X-ray luminosity of $22.3\times 10^{43}$~erg~s$^{-1}$, $\Gamma=1.61$ and  radio luminosity at $408$~MHz of  $83.5\times 10^{42}$~erg~s$^{-1}$.

\begin{figure}
\centering
\includegraphics[width=0.47\textwidth]{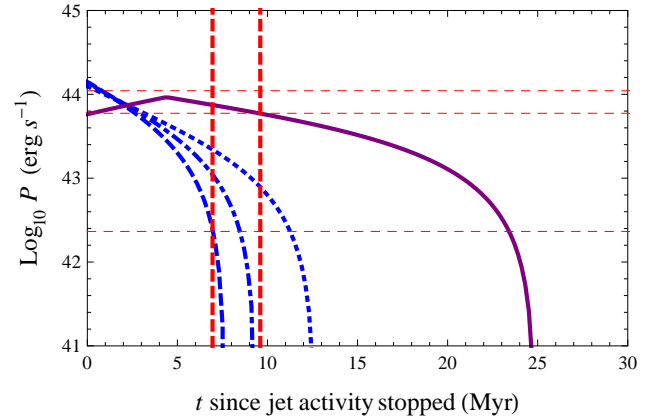}
\caption{Evolution of X-ray and radio luminosities of model [14] for HDF\,130 after jet activity has halted. The solid track shows the $1$~keV X-ray power. The dashed/dot-dashed/dotted blue lines show the evolution of the $240$/$151$/$74$~MHz luminosities. Vertical dashed lines show the duration in the evolution of the source that is similar to HDF\,130. The horizontal thin dashed lines show the 
$\pm 30$~per~cent range of acceptable X-ray luminosities and the
$3\sigma$ $240$~MHz luminosity upper limit of emission.}
\label{fig1}
\end{figure}

We consider a range of parameters and test which models (at some point in their evolution, the congruent observational window, denoted by $t_{\rm obs}$) give a lobe length, axial ratio, X-ray luminosity and photon index that agree with HDF\,130 to within $30$~per~cent as well as a $240$~MHz radio luminosity that is below the $3\sigma$ flux density limit (this percentage is arbitrary but was chosen to be large to search for sets of parameters that model sources similar to the observations without having to narrow down any parameter too precisely). We also test which models agree with 6C\,0905+3955 to within $60$~per~cent (a higher margin of error is considered for this source as it is asymmetric which our model does not account for).
We look at jet energies ranging from $5\times10^{37}$~W to $10^{40}$~W, jet lifetimes from 
$10^5$~years to $10^9$~years, $\beta=1.5$ and $2$, $\rho_0=1.67\times 10^{-23}, 1.67\times 10^{-22}$~kg m$^{-3}$,
$\gamma_{\rm min}$ of $1$ , $10^3$ and  $10^4$, and injection indices of $2.14,2.5$ and $3$.
The parameters considered are listed in Table~\ref{tblpar}. 
A total of $2304$ models were tested by considering all combinations of the parameters in Table~\ref{tblpar}. Of these $2304$ runs, $19$ end up closely resembling all five of the observational features of HDF\,130 at some point in the evolution of the source. These models are presented in Table~\ref{tbl1}. 
The models that closely describe 6C\,0905+3955 are presented in Table~\ref{tbl1b}.
The predicted observable features of the congruent models are presented in Tables~\ref{tbl2} and \ref{tbl2b} of the Appendix, and are compared to the observed values.
Figure~\ref{fig1} shows an example (model [14] for HDF\,130) of how X-ray luminosities and radio luminosities are predicted by the model to evolve with time.

We may obtain a constraint on $p$ from $\Gamma$ for an active source. The value of $p$ is probably steeper than the value implied by $\Gamma$ (namely, $p=2\Gamma-1$) to reflect the empirical point that for an active classical double radio source the spectrum has a gradient flatter at lower $\gamma$ (which may reflect a cutoff/turnover/$\gamma_{\rm min}$) and not representative of the high $\gamma$ particles responsible for GHz emission. Table~\ref{tbl1b} for the source 6C\,0905+3955 shows the values of $p$ greater than the value implied by $\Gamma$ in bold. 
For a source that has turned off, such as HDF\,130, the observed $\Gamma$ grows with time regardless of the value of $p$ for the injection spectrum, so we cannot make such a constraint.

\begin{table}
\centering
\caption{Models that agree with HDF\,130 observations}\label{tbl1}
\begin{tabular}{@{}lccccccc@{}}
\hline\hline
\noalign{\smallskip}
\# & $Q_{\rm j}^{\rm a}$ & $t_{\rm j}^{\rm b}$ & $\gamma_{\rm min}$ & $p$ & $\beta$ & $\rho_0^{\rm c}$ & $t_{\rm obs}^{\rm d}$  \\[2pt]
\hline 
\noalign{\smallskip}
[1] & $1$ & $100$ & $1000$ & $3$ & $1.5$ & $16.7$ & $1.07$ to $1.11$\\[2pt]
[2] & $2$ & $100$ & $1000$ & $3$ & $2$ & $16.7$ & $1.05$ to $1.11$\\[2pt]
[3] & $5$ & $50$ & $1000$ & $2.5$ & $1.5$ & $1.67$ & $1.13$ to $1.29$\\[2pt]
[4] & $5$ & $50$ & $1000$ & $2.5$ & $1.5$ & $16.7$ & $1.16$ to $1.27$\\[2pt]
[5] & $5$ & $50$ & $1000$ & $3$ & $1.5$ & $1.67$ & $1.11$ to $1.31$\\[2pt]
[6] & $5$ & $50$ & $1000$ & $3$ & $2$ & $16.7$ & $1.18$ to $1.27$\\[2pt]
[7] & $5$ & $100$ & $1000$ & $2.14$ & $1.5$ & $16.7$ & $1.09$ to $1.11$\\[2pt]
[8] & $5$ & $100$ & $1000$ & $2.5$ & $1.5$ & $16.7$ & $1.10$ to $1.15$\\[2pt]
[9] & $10$ & $50$ & $1000$ & $2.14$ & $1.5$ & $16.7$ & $1.16$ to $1.20$\\[2pt]
[10] & $10$ & $50$ & $1000$ & $2.14$ & $2$ & $16.7$ & $1.17$ to $1.31$\\[2pt]
[11] & $10$ & $50$ & $1000$ & $2.5$ & $1.5$ & $16.7$ & $1.16$ to $1.20$\\[2pt]
[12] & $10$ & $50$ & $1000$ & $2.5$ & $2$ & $16.7$ & $1.25$ to $1.33$\\[2pt]
[13] & $10$ & $100$ & $1000$ & $2.14$ & $1.5$ & $16.7$ & $1.10$ to $1.13$\\[2pt]
[14] & $20$ & $10$ & $1000$ & $2.5$ & $2$ & $1.67$ & $1.70$ to $1.95$\\[2pt]
[15] & $20$ & $10$ & $1000$ & $3$ & $2$ & $1.67$ & $1.68$ to $2.04$\\[2pt]
[16] & $20$ & $50$ & $1000$ & $2.14$ & $1.5$ & $16.7$ & $1.18$ to $1.22$\\[2pt]
[17] & $50$ & $50$ & $1$ & $2.14$ & $1.5$ & $16.7$ & $1.15$ to $1.18$\\[2pt]
[18] & $100$ & $5$ & $1000$ & $2.14$ & $2$ & $1.67$ & $2.45$ to $2.53$\\[2pt]
[19] & $100$ & $5$ & $1000$ & $2.5$ & $2$ & $1.67$ & $2.45$ to $2.49$\\[2pt]
\hline \noalign{\smallskip}
\multicolumn{8}{l}{$^{\rm a}$ ($\times10^{38}$ W) \,\,\, 
$^{\rm b}$ (Myr) \,\,\,
$^{\rm c}$ ($\times10^{-23}~{\rm kg}~{\rm m}^{-3}$)} \\
$^{\rm d}$ ($\times t_{\rm j}$) \\[2pt]
\end{tabular}
\end{table}

\begin{table}
\centering
\caption{Models that agree with 6C\,0905+3955 observations}\label{tbl1b}
\begin{tabular}{@{}lcccccc@{}}
\hline\hline
\noalign{\smallskip}
\# & $Q_{\rm j}^{\rm a}$ & $\gamma_{\rm min}$ & $p$ & $\beta$ & $\rho_0^{\rm b}$ & $t_{\rm obs}^{\rm c}$  \\[2pt]
\hline 
\noalign{\smallskip}
[1] & $50$ & $1$ & $2.14$ & $1.5$ & $1.67$ & $0.42$ to $0.52$\\[2pt]
[2] & $50$ & $1$ & $2.14$ & $2$ & $1.67$ & $0.54$ to $0.60$\\[2pt]
[3] & $50$ & $1000$ & $2.14$ & $1.5$ & $1.67$ & $0.45$ to $0.52$\\[2pt]
[4] & $50$  & $1000$ & $2.14$ & $2$ & $1.67$ & $0.21$ to $0.28$\\[2pt]
[5] & $50$  & $1000$ & {\bf 2.5} & $2$ & $1.67$ & $0.17$ to $0.28$\\[2pt]  
[6] & $50$  & $1000$ & {\bf 3} & $2$ & $1.67$ & $0.17$ to $0.28$\\[2pt]    
[7] & $100$  & $1$ & $2.14$ & $1.5$ & $1.67$ & $0.30$ to $0.41$\\[2pt] 
[8] & $100$  & $1$ & $2.14$ & $2$ & $1.67$ & $0.21$ to $0.22$\\[2pt] 
[9] & $100$  & $1$ & $2.14$ & $2$ & $16.7$ & $0.42$ to $0.47$\\[2pt] 
[10] & $100$  & $1000$ & $2.14$ & $2$ & $1.67$ & $0.21$ to $0.22$\\[2pt] 
\hline \noalign{\smallskip}
\multicolumn{7}{l}{$^{\rm a}$ ($\times10^{38}$ W) \,\,\, 
$^{\rm b}$ ($\times10^{-23}~{\rm kg}~{\rm m}^{-3}$)}  \,\,\,
$^{\rm c}$ ($\times 100$~Myr) \\[2pt]
\end{tabular}
\end{table}

The models that do reasonably well at describing 
a source similar to  HDF\,130 all suggest that the source is being viewed after the jets have turned off (by at least $5$~Myr) while the radio lobe emission is falling rapidly or has already dropped to below telescope sensitivity. All models except one (which has a short observational window) require $\gamma_{\rm min}$ to be $1000$ rather than $1$. With $\gamma_{\rm min}=1$ the FR~II object in the models are not bright enough in the X-rays 
(see \S~\ref{sec:gmin} 
for a discussion on the effects of varying $\gamma_{\rm min}$)
during the times when the other parameters (length, photon index, etc.) agree within $\pm30$~per~cent error to what is observed for HDF\,130. The models suggest that the jet lifetime of HDF\,130 is on the order of $1$--$100$~Myr, and that the source was powerful, with jet power on the order of $10^{38}$--$10^{40}$~W. The window in the time evolution of the source during which it is consistent with HDF\,130 in the different models is typically a few Myr.

The models imply a long jet lifetime of at least $20$~Myr for 6C\,0905+3955, but we can only place a lower limit for jet lifetime for an active jet. The jet power is strong: $5\times10^{39}$--$10^{40}$~W. A $\gamma_{\rm min}$ of $1000$ (or, briefly, $1$) are preferred over $\gamma_{\rm min}=10^4$ to accurately describe the lobe luminosity in both the X-ray and radio simultaneously. Likely $\gamma_{\rm min}$ is similar to or somewhat higher than $1000$, as no X-ray emission (other than the synchrotron extrapolation observed by \cite{2008MNRAS.386.1774E} with XMM) is observed in the hotspots of the source. The models under-predict the unusually large axial ratio of 6C\,0905+3955 and this may be due to the fact that we averaged the two asymmetric arm lengths.

\subsection{The effects of $\gamma_{\rm min}$}\label{sec:gmin}

\begin{figure}
\centering
\includegraphics[width=0.47\textwidth]{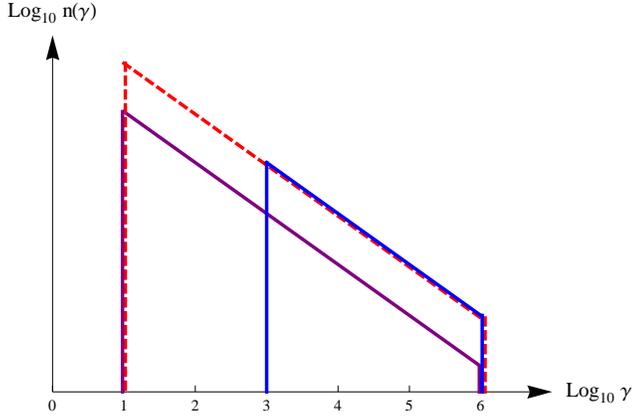}
\caption{
The effect of $\gamma_{\rm min}$ on the injection spectrum.
Increasing the parameter $\gamma_{\rm min}$ alone while conserving the electron energy density in our model (which is linked to the pressure at time of injection into the lobe) results in more higher $\gamma$ particles (see the two solid color injection spectra), making the source
brighter. This is because the electron energy density is determined by a minimum energy argument and is linked
to the lobe pressure, which is determined from the jet power and the environmental parameters.
Extending a spectrum to include lower $\gamma$ particles (dashed spectrum), which may seem innocuous because
these particles do not contribute to the $1{\rm keV}$ X-ray and $151{\rm MHz}$ radio emission, increases
the energy density of electrons, hence the pressure, and 
alters the growth of the lobes.
}
\label{fig:gmin}
\end{figure}

The question arises why changing $\gamma_{\rm min}$ has an effect on the evolution of the source because 
the electrons below $\gamma_{\rm min}$ do not contribute to the $1{\rm keV}$ X-ray and $151{\rm MHz}$ radio emission.
In short, increasing $\gamma_{\rm min}$ only from a set of given parameters will drive the source to become
brighter without changing lobe growth because the electron energy density of the injection spectrum 
in our model stays the same as it is assumed to be linked, by a minimum energy argument, to the lobe pressure by a constant factor, and
the pressure is determined by the jet power and the environmental parameters.
Having the injected electron energy density kept constant but
increasing $\gamma_{\rm min}$ yields more higher $\gamma$ particles (see the two solid color injection spectra 
in Figure~\ref{fig:gmin}) and hence a brighter source at $1{\rm keV}$ and $151{\rm MHz}$.
Alternatively, we can ask ourselves what is the effect of extending the electron energy spectrum below $\gamma_{\rm min}$, without keeping electron energy density constant. If we assume an injected electron spectrum with high $\gamma_{\rm min}$
and then extrapolate the spectrum to include lower $\gamma$ particles (see the dashed injection spectrum in Figure~\ref{fig:gmin}) then
this will increase the energy density of electrons, hence the pressure
and the dynamics of the source are altered and the lobes grow much larger. The combination of lobe size and source brightness
will constrain possible values of $\gamma_{\rm min}$ in our sources.

\section{Discussion and Conclusion}\label{sec:Discussion}

We matched the observational features of HDF\,130 and 6C\,0905+3955 to analytic models of the evolution of the lobe length, axial ratio, photon index, and X-ray and radio luminosities of an FR~II object. {\it Only periods of time when the source is no longer active in the models are congruent to the observations of HDF\,130, supporting the idea that HDF\,130 is an IC ghost of a giant radio source.} The models suggest HDF\,130 experienced jet activity for a period of around $5$--$100$~Myr, and that we are viewing the object at least a few Myr after the jets have turned off, which is why the source is not bright in the radio.
6C\,0905+3955 is inferred to have had an active jet for at least $20$~Myr and may have a slightly higher intrinsic jet power than HDF\,130. 

Predicted radio luminosities at lower frequencies ($151$ and $74$ MHz) for the models that agree with HDF\,130 are included in Table~\ref{tbl2}. Even at $151$~MHz the ghost source may not be observable. However, the models predict that the source will be observable in the $74$~MHz band, with luminosity on the order of $10^{43}$~erg~s$^{-1}$, 
or, equivalently, $10^{20}$~W~Hz$^{-1}$~sr$^{-1}$.
New low-frequency observations of HDF\,130 could test this.

Some models predict the lobe lengths in the lower limit of the error tolerance during the observational congruent window of HDF\,130, while predicting the other features accurately. If such is the case, perhaps the surrounding density profile is not as simple as we have assumed it to be and allows for the lobes to grow larger than in our models while staying bright in the X-ray. It is plausible that the source is expanding into a pre-existing lobe from a previous episode of jet activity, which cleared away some of the surrounding material and would mean expansion losses are smaller and the lobes can grow larger and brighter.

Importantly, the minimum Lorentz factor of injected particles into the lobe for HDF\,130 is found to be on the order of $\gamma_{\rm min}=1000$ rather than $\gamma_{\rm min}=1$. 
Even a $\gamma_{\rm min}=30$ or $\gamma_{\rm min}=100$ is not preferred by HDF\,130: repeating fitting the observable properties of HDF\,130 with models that have
$\gamma_{\rm min}=30$ and $\gamma_{\rm min}=100$ gives only $4$ congruent models observational windows of at most $3$~Myr, reported in Table~\ref{tbl1c}. 

In the model, a higher $\gamma_{\rm min}$ (while keeping injected electron energy density constant) will produce brighter sources without affecting the lobe growth, which is determined by the jet power and the surrounding density profile. Increasing the jet power makes the jet grow larger and brighter. It is the combination of  HDF\,130 lobe size, which is not exceptionally large, and X-ray brightness which forces the model to require $\gamma_{\rm min}\sim 1000$ to agree with the observational features.

\begin{table}
\centering

\caption{Additional models investigated that agree with HDF\,130 observations}\label{tbl1c}
\begin{tabular}{@{}lccccccc@{}}
\hline\hline
\noalign{\smallskip}
\# & $Q_{\rm j}^{\rm a}$ & $t_{\rm j}^{\rm b}$ & $\gamma_{\rm min}$ & $p$ & $\beta$ & $\rho_0^{\rm c}$ & $t_{\rm obs}$ ($\times t_{\rm j}$) \\[2pt]
\hline 
\noalign{\smallskip}
[1] & $50$ & $50$ & $30$ & $2.14$ & $1.5$ & $16.7$ & $1.16$ to $1.21$\\[2pt]
[2] & $10$ & $100$ & $100$ & $2.14$ & $1.5$ & $16.7$ & $1.09$ to $1.12$\\[2pt]
[3] & $20$ & $50$ & $100$ & $2.14$ & $1.5$ & $16.7$ & $1.16$ to $1.20$\\[2pt]
[4] & $50$ & $50$ & $100$ & $2.5$ & $1.5$ & $16.7$ & $1.15$ to $1.18$\\[2pt]
\hline \noalign{\smallskip}
\multicolumn{8}{l}{$^{\rm a}$ ($\times10^{38}$ W) \,\,\, 
$^{\rm b}$ (Myr) \,\,\,
$^{\rm c}$ ($\times10^{-23}~{\rm kg}~{\rm m}^{-3}$)}  \\[2pt]
\end{tabular}
\end{table}
The minimum injected Lorentz factor for 6C\,0905+3955 is also found to be most likely on the order of $\gamma_{\rm min}=1000$ (or also marginally $1$), based on only best matching the total lobe luminosities predicted by the model to the observed luminosities. 
Considering only the models where $p$ is steeper than implied by $\Gamma$ (these values are bold in Table~\ref{tbl1b}), we see that only $\gamma_{\rm min}=1000$ is preferred.
In previous observations, no X-ray emission is seen in a hotspot of 6C\,0905+3955 (other than highly energetic X-ray synchrotron requiring extremely high Lorentz factors  \cite{2008MNRAS.386.1774E}) which suggests that there is a low-energy cutoff of the freshly injected particles into the lobe above the $\gamma\sim10^3$ particles required for X-ray emission from upscattering on the CMB. Likely, the minimum energy cutoff is just above the critical Lorentz factor which would result in X-ray emission from the hotspot. It is important to note that 6C\,0905+3955 may be more complicated than described by our simple model, because 6C\,0905+3955 is asymmetric, probably due to an asymmetric surrounding environment. There may also be complex mechanisms happening in the lobes, such as reflected shocks or interruptions of the jet at the hotspot \citep{1995MNRAS.277..995L}, which are so far unaccounted for by our model.

The chosen value of $\gamma_{\rm min}$ varies by orders of magnitude in previous papers, as it often has to be estimated.
The minimum Lorentz factor is assumed to be typically $1$ in previous models of FR~II evolution by \cite{1997MNRAS.292..723K}, \cite{1997MNRAS.286..215K}, \cite{1999AJ....117..677B} and \cite{2010MNRAS.407.1998N}.
\cite{2005ApJ...626..733C} use a value of $10$, \cite{1991ApJ...383..554C} use a value of $100$, and \cite{1998Natur.395..457W} use a value of $1000$.
If HDF\,130 and 6C\,0905+3955 are typical sources, it may be the case that the minimum energy of particles injected into the lobes is large. The value of $\gamma_{\rm min}$ may at first appear as an eclectic, unimportant detail, but \citet{2011MNRAS.413.1107M} show that the typical value of $\gamma_{\rm min}$ can significantly affect estimates for the total population of FR~II sources from a radio luminosity function as it changes the time sources fall below a given flux limit in their evolution. A higher $\gamma_{\rm min}$ will also increase the detectability of IC ghosts.

\section*{Acknowledgments}
PM would like to acknowledge the award of a Weissman grant from Harvard University. KMB and ACF thank the Royal Society for support.

\bibliography{mybib}{}

\appendix
\section[]{Model predictions}

\begin{table*}
\centering
\begin{minipage}{168mm}
\caption{Model predictions during congruent observational window compared to HDF\,130 observations}\label{tbl2}
\begin{tabular}{@{}lccccccc@{}}
\hline\hline
\noalign{\smallskip}
\# &  lobe length \footnote{(kpc)} & axial ratio 
& $L_{{\rm x},1{\rm keV}}$ \footnote{($\times10^{43}$ erg s$^{-1}$)} 
& $\Gamma$ & $L_{{\rm r},240{\rm MHz}}$ \footnote{$(\times10^{42}$ erg s$^{-1}$)} 
& $L_{{\rm r},151{\rm MHz}}$ \footnote{$(\times10^{42}$ erg s$^{-1}$)}
& $L_{{\rm r},74{\rm MHz}}$ \footnote{$(\times10^{42}$ erg s$^{-1}$)} \\[2pt]
\hline 
\noalign{\smallskip}
obs. & $345$                             & $2.0$              & $8.5$              & $2.65$              & $<2.3$  & & \\[2pt]

[1] & $278$ to $281$ & $2.1$ to $2.1$ & $7.8$ to $6.2$ & $2.6$ to $3.1$ & $1.3$ to $0$ & $3.8$ to $0.1$ & $10.5$ to $3.5$ \\[2pt]

& ($-0.2$ to $-0.19$)\footnote{fractional deviation} & ($0.05$ to $0.05$) & ($-0.08$ to $-0.27$) & ($-0.03$ to $0.18$) & ($1.62$ to $0$)\footnote{fraction of $1\sigma$ upper limit} & & \\[2pt]

[2] & $435$ to $444$ & $2.5$ to $2.5$ & $6.2$ to $6.1$ & $2.0$ to $2.9$ & $0.5$ to $0$ & $1.0$ to $0$ & $2.3$ to $0.3$ \\[2pt]

& ($0.26$ to $0.29$) & ($0.25$ to $0.25$) & ($-0.27$ to $-0.28$) & ($-0.25$ to $0.09$) & ($0.63$ to $0$) & & \\[2pt]

[3] & $391$ to $408$ & $2.6$ to $2.6$ & $7.8$ to $6.0$ & $1.9$ to $2.7$ & $1.8$ to $0$ & $3.8$ to $0$ & $7.1$ to $0.1$ \\[2pt]

& ($0.13$ to $0.18$) & ($0.3$ to $0.3$) & ($-0.08$ to $-0.29$) & ($-0.27$ to $0.03$) & ($2.32$ to $0$) & & \\[2pt]

[4] & $306$ to $317$ & $2.4$ to $2.4$ & $8.7$ to $5.9$ & $2.3$ to $2.8$ & $1.5$ to $0$ & $5.8$ to $0$ & $13.3$ to $2.0$ \\[2pt]

& ($-0.11$ to $-0.08$) & ($0.18$ to $0.18$) & ($0.03$ to $-0.3$) & ($-0.12$ to $0.07$) & ($1.89$ to $0$) & & \\[2pt]

[5] & $389$ to $410$ & $2.6$ to $2.6$ & $10.7$ to $6.5$ & $2.0$ to $3.4$ & $1.7$ to $0$ & $3.7$ to $0$ & $8.0$ to $0$ \\[2pt]

& ($0.13$ to $0.19$) & ($0.3$ to $0.3$) & ($0.26$ to $-0.23$) & ($-0.24$ to $0.29$) & ($2.25$ to $0$) & & \\[2pt]

[6] & $308$ to $317$ & $2.4$ to $2.4$ & $10.9$ to $6.8$ & $2.8$ to $3.4$ & $0$ to $0$ & $1.6$ to $0$ & $9.0$ to $0.9$ \\[2pt]

& ($-0.11$ to $-0.08$) & ($0.18$ to $0.18$) & ($0.28$ to $-0.2$) & ($0.07$ to $0.29$) & ($0.01$ to $0$) & & \\[2pt]

[7] & $363$ to $365$ & $2.3$ to $2.3$ & $6.8$ to $6.1$ & $2.2$ to $2.4$ & $1.0$ to $0$ & $6.8$ to $1.6$ & $14.2$ to $10.0$ \\[2pt]

& ($0.05$ to $0.06$) & ($0.16$ to $0.16$) & ($-0.2$ to $-0.28$) & ($-0.15$ to $-0.11$) & ($1.3$ to $0$) & & \\[2pt]

[8] & $364$ to $369$ & $2.3$ to $2.3$ & $10.3$ to $6.9$ & $2.6$ to $3.3$ & $0$ to $0$ & $2.9$ to $0$ & $13.9$ to $1.7$ \\[2pt]

& ($0.05$ to $0.07$) & ($0.16$ to $0.16$) & ($0.21$ to $-0.19$) & ($-0.01$ to $0.23$) & ($0$ to $0$) & & \\[2pt]

[9] & $249$ to $252$ & $2.2$ to $2.2$ & $7.3$ to $6.1$ & $2.4$ to $2.8$ & $0.4$ to $0$ & $24.4$ to $0.7$ & $54.2$ to $34.6$ \\[2pt]

& ($-0.28$ to $-0.27$) & ($0.09$ to $0.09$) & ($-0.13$ to $-0.28$) & ($-0.09$ to $0.04$) & ($0.56$ to $0$) & & \\[2pt]

[10] & $387$ to $404$ & $2.6$ to $2.6$ & $8.6$ to $6.0$ & $2.1$ to $2.7$ & $2.0$ to $0$ & $9.0$ to $0$ & $16.9$ to $0.3$ \\[2pt]

& ($0.12$ to $0.17$) & ($0.29$ to $0.29$) & ($0.01$ to $-0.29$) & ($-0.22$ to $0$) & ($2.6$ to $0$) & & \\[2pt]

[11] & $249$ to $252$ & $2.2$ to $2.2$ & $10.9$ to $8.3$ & $2.8$ to $3.3$ & $0$ to $0$ & $19.3$ to $0$ & $68.2$ to $36.7$ \\[2pt]

& ($-0.28$ to $-0.27$) & ($0.09$ to $0.09$) & ($0.28$ to $-0.02$) & ($0.07$ to $0.26$) & ($0$ to $0$) & & \\[2pt]

[12] & $397$ to $406$ & $2.6$ to $2.6$ & $10.8$ to $7.6$ & $2.7$ to $3.4$ & $0$ to $0$ & $0$ to $0$ & $5.8$ to $0$ \\[2pt]

& ($0.15$ to $0.18$) & ($0.29$ to $0.29$) & ($0.27$ to $-0.1$) & ($0.02$ to $0.28$) & ($0$ to $0$) & & \\[2pt]

[13] & $444$ to $447$ & $2.5$ to $2.5$ & $10.8$ to $9.1$ & $2.3$ to $2.6$ & $0$ to $0$ & $7.6$ to $0$ & $22.4$ to $11.2$ \\[2pt]

& ($0.29$ to $0.3$) & ($0.25$ to $0.25$) & ($0.27$ to $0.07$) & ($-0.13$ to $-0.03$) & ($0$ to $0$) & & \\[2pt]

[14] & $243$ to $257$ & $2.5$ to $2.5$ & $7.4$ to $6.0$ & $2.1$ to $2.4$ & $2.2$ to $0$ & $10.7$ to $0$ & $21.6$ to $8.2$ \\[2pt]

& ($-0.3$ to $-0.26$) & ($0.25$ to $0.25$) & ($-0.12$ to $-0.3$) & ($-0.19$ to $-0.11$) & ($2.84$ to $0$) & & \\[2pt]

[15] & $242$ to $262$ & $2.5$ to $2.5$ & $9.9$ to $6.0$ & $2.6$ to $3.1$ & $0.9$ to $0$ & $6.6$ to $0$ & $20.1$ to $2.4$ \\[2pt]

& ($-0.3$ to $-0.24$) & ($0.25$ to $0.25$) & ($0.16$ to $-0.3$) & ($-0.03$ to $0.17$) & ($1.19$ to $0$) & & \\[2pt]

[16] & $305$ to $309$ & $2.3$ to $2.3$ & $10.7$ to $8.5$ & $2.7$ to $3.3$ & $0$ to $0$ & $16.1$ to $0$ & $76.6$ to $41.9$ \\[2pt]

& ($-0.12$ to $-0.11$) & ($0.17$ to $0.17$) & ($0.26$ to $0$) & ($0$ to $0.23$) & ($0$ to $0$) & & \\[2pt]

[17] & $393$ to $397$ & $2.6$ to $2.6$ & $7.3$ to $6.2$ & $2.5$ to $2.8$ & $0.5$ to $0$ & $31.6$ to $6.2$ & $70.4$ to $50.1$ \\[2pt]

& ($0.14$ to $0.15$) & ($0.3$ to $0.3$) & ($-0.14$ to $-0.27$) & ($-0.05$ to $0.07$) & ($0.68$ to $0$) & & \\[2pt]

[18] & $242$ to $245$ & $2.6$ to $2.6$ & $6.3$ to $6.0$ & $2.5$ to $2.7$ & $0$ to $0$ & $0$ to $0$ & $31.9$ to $23.9$ \\[2pt]

& ($-0.3$ to $-0.29$) & ($0.29$ to $0.29$) & ($-0.26$ to $-0.3$) & ($-0.04$ to $0.02$) & ($0$ to $0$) & & \\[2pt]

[19] & $242$ to $243$ & $2.6$ to $2.6$ & $7.7$ to $7.4$ & $3.3$ to $3.4$ & $0$ to $0$ & $0$ to $0$ & $25.6$ to $20.9$ \\[2pt]

& ($-0.3$ to $-0.29$) & ($0.29$ to $0.29$) & ($-0.09$ to $-0.13$) & ($0.25$ to $0.29$) & ($0$ to $0$) & & \\[2pt]
\hline
\end{tabular}
\end{minipage}
\end{table*}

\begin{table*}
\begin{minipage}{168mm}
\centering
\caption{Model predictions during congruent observational window compared to 6C\,0905+3955 observations}\label{tbl2b}
\begin{tabular}{@{}lccccccc@{}}
\hline\hline
\noalign{\smallskip}
\# &  lobe length \footnote{(kpc)} & axial ratio 
& $L_{{\rm x},1{\rm keV}}$ \footnote{($\times10^{43}$ erg s$^{-1}$)} 
& $\Gamma$ & $L_{{\rm r},408{\rm MHz}}$ \footnote{$(\times10^{42}$ erg s$^{-1}$)} 
 \\[2pt]
\hline 
\noalign{\smallskip}
obs. & $472$                             & $8.0$              & $22.3$              & $1.61$              & $83.5$  & & \\[2pt]

[1] & $626$ to $752$ & $3.2$ to $3.3$ & $10.2$ to $11.2$ & $1.8$ to $1.8$ & $46.5$ to $34.0$  \\[2pt]

& ($0.33$ to $0.59$) & ($-0.6$ to $-0.58$) & ($-0.54$ to $-0.5$) & ($0.1$ to $0.11$) & ($-0.45$ to $-0.6$) \\[2pt]

[2] & $675$ to $750$ & $3.2$ to $3.3$ & $11.6$ to $12.0$ & $1.8$ to $1.8$ & $50.1$ to $41.0$  \\[2pt]

& ($0.43$ to $0.59$) & ($-0.6$ to $-0.59$) & ($-0.48$ to $-0.46$) & ($0.12$ to $0.13$) & ($-0.4$ to $-0.51$) \\[2pt]

[3] & $664$ to $752$ & $3.2$ to $3.3$ & $19.8$ to $21.3$ & $1.0$ to $1.1$ & $131.1$ to $105.7$  \\[2pt]

& ($0.41$ to $0.59$) & ($-0.59$ to $-0.58$) & ($-0.11$ to $-0.04$) & ($-0.35$ to $-0.32$) & ($0.56$ to $0.26$) \\[2pt]

[4] & $566$ to $754$ & $3.4$ to $3.6$ & $9.0$ to $10.7$ & $0.8$ to $0.8$ & $77.9$ to $46.0$  \\[2pt]

& ($0.2$ to $0.6$) & ($-0.58$ to $-0.55$) & ($-0.6$ to $-0.52$) & ($-0.53$ to $-0.48$) & ($-0.07$ to $-0.45$) \\[2pt]

[5] & $458$ to $754$ & $3.2$ to $3.6$ & $14.2$ to $19.0$ & $0.7$ to $0.9$ & $131.0$ to $49.8$  \\[2pt]

& ($-0.03$ to $0.6$) & ($-0.6$ to $-0.55$) & ($-0.36$ to $-0.15$) & ($-0.54$ to $-0.46$) & ($0.56$ to $-0.41$) \\[2pt]

[6] & $458$ to $754$ & $3.2$ to $3.6$ & $21.0$ to $27.7$ & $0.8$ to $0.9$ & $112.8$ to $37.5$  \\[2pt]

& ($-0.03$ to $0.6$) & ($-0.6$ to $-0.55$) & ($-0.06$ to $0.24$) & ($-0.52$ to $-0.44$) & ($0.34$ to $-0.55$) \\[2pt]

[7] & $572$ to $748$ & $3.2$ to $3.4$ & $14.6$ to $17.2$ & $1.7$ to $1.8$ & $132.4$ to $89.4$  \\[2pt]

& ($0.21$ to $0.58$) & ($-0.6$ to $-0.57$) & ($-0.34$ to $-0.23$) & ($0.08$ to $0.1$) & ($0.58$ to $0.06$) \\[2pt]

[8] & $713$ to $747$ & $3.7$ to $3.7$ & $9.$ to $9.2$ & $1.6$ to $1.6$ & $42.2$ to $38.9$  \\[2pt]

& ($0.51$ to $0.58$) & ($-0.54$ to $-0.53$) & ($-0.6$ to $-0.59$) & ($0.02$ to $0.02$) & ($-0.5$ to $-0.54$) \\[2pt]

[9] & $662$ to $741$ & $3.3$ to $3.4$ & $17.4$ to $18.4$ & $1.8$ to $1.8$ & $130.7$ to $108.2$  \\[2pt]

& ($0.4$ to $0.57$) & ($-0.59$ to $-0.58$) & ($-0.22$ to $-0.18$) & ($0.11$ to $0.12$) & ($0.56$ to $0.29$) \\[2pt]

[10]& $713$ to $747$ & $3.7$ to $3.7$ & $15.1$ to $15.5$ & $0.8$ to $0.8$ & $131.1$ to $121.$  \\[2pt]

& ($0.51$ to $0.58$) & ($-0.54$ to $-0.53$) & ($-0.32$ to $-0.3$) & ($-0.53$ to $-0.52$) & ($0.56$ to $0.44$) \\[2pt]

\hline
\end{tabular}
\end{minipage}
\end{table*}

\bsp
\label{lastpage}
\end{document}